\documentclass[%
 reprint,
 amsmath,amssymb,
 aps,
]{revtex4-2}

\usepackage{graphicx}
\usepackage{dcolumn}
\usepackage{bm}
\usepackage{xcolor}
\usepackage{comment}

\begin{document}

\preprint{APS/123-QED}

\title{Fokker-Planck entropic force interpretation of galactic rotation curves}

\author{V.~S. Morales-Salgado}
\affiliation{CEDIP-C\'amara de Diputados, H. Congreso de la Uni\'on 66, El Parque, Venustiano Carranza, 15960, Ciudad de M\'exico, M\'exico}
\affiliation{Asociación Mexicana para el Avance de la Ciencia (AMEXAC), Ciudad de México, México}
\author{H. Mart\'inez-Huerta }%
\email{humberto.martinezhuerta@udem.edu}
\affiliation{Departamento de Física y Matemáticas, Universidad de Monterrey,\\Av. Morones Prieto 4500, 66238, San Pedro Garza Garc\'ia NL, México 
}%
\author{P.~I. Ram\'irez-Baca}
\altaffiliation[Also at ]{School of Engineering and Sciences, Tecnologico de Monterrey, Atizapan 52926, Mexico}
\altaffiliation[partial work was carried in ]{Facultad de Ciencias, Universidad Aut\'onoma de San Luis Potos\'i Campus Pedregal, Av. Parque Chapultepec 1610, Col. Privadas del Pedregal, San Luis Potos\'i, SLP, 78217, Mexico.}
\affiliation{Departamento de Física y Matemáticas, Universidad de Monterrey,\\Av. Morones Prieto 4500, 66238, San Pedro Garza Garc\'ia NL, M\'exico.}

\date{April 22, 2026}

\begin{abstract}

We investigate whether the discrepancy between observed galactic rotation curves and those predicted from baryonic matter can be interpreted as the manifestation of an emergent entropic force. Starting from a minimal statistical framework, we derive an effective radial force from a stationary solution of the Fokker–Planck equation under simple and physically motivated assumptions. 
We confront this Fokker–Planck entropic (FPE) model with high-quality rotation curves from the SPARC database, performing a systematic comparison with standard halo profiles, including Navarro–Frenk–White (NFW), Burkert, and pseudo-isothermal (ISO) models. The FPE model provides fits of comparable or improved statistical quality than traditional profiles, while yielding stellar mass-to-light ratios within physically consistent ranges, in contrast to NFW and Burkert fits that often approach prior limits.
Beyond reproducing rotation curves, the model naturally gives rise to strong correlations between its characteristic parameter and global galaxy properties, including the flat rotation velocity and infrared luminosity. These relations are consistent with well-known empirical scaling laws such as the Tully–Fisher relation, suggesting that the proposed framework captures key aspects of the underlying dynamics.
Our results indicate that a minimal entropic-force description, grounded in statistical mechanics, can account for galactic rotation curves while simultaneously encoding their scaling relations, offering a complementary and physically motivated perspective to standard dark matter halo interpretations of galaxy rotation curves.

\end{abstract}

\maketitle

Current models framing astronomical observations demand the presence of exotic components introduced to explain the difference between how objects in the sky ought to move, according to some preconceived notion, and how they are actually observed to move \citep{s14}.
However, as has been coined colloquially, the dark sector of the universe still eludes detection in a plethora of experiments and has no clear origin within the standard model of fundamental particles~\citep{baudis2016dark}.
This compels us to consider alternative explanations for the observed motion of celestial bodies.

One of the earliest motivations to introduce the concept of dark matter is the so-called {\it galactic rotation curves problem}.
A rotation curve describes the rotational velocity of a test mass as a function of the radial distance to the galactic center.
Starting with the seminal work of  \citet{Rubin:1970zza}, numerous studies have been performed to obtain rotation curves of a vast sample of galaxies. 
However, there exist discrepancies between astrophysical observations and rotational velocities modeled according to predicted mass profiles. 
This is because, while a classical macroscopic treatment of gravity predicts a fall of orbital speeds like $r^{-1/2}$ at large radii, observations show that velocity remains constant with respect to distance far from galactic centers \citep{Persic:1995ru}.

In this work, we use observations compiled by the Spitzer Photometry and Accurate Rotation Curves (SPARC) project. 
This is a major database of high-quality rotation curves spanning a broad range of morphologies, luminosities, and surface brightnesses \citep{Lelli:2016zqa}. 
Studies using this data set with a variety of models can be found in literature, see for instance~\cite{galaxies9010017,2018MNRAS.4,Hernandez-Almada:2017mtm, Li:2019ksm, 2018JCAP, 2018yao}, although the true nature of the discrepancies remains unsolved. 

Most proposals to solve the galaxy rotation curve problem, including dark matter, take the general approach of adding an extra term $F_E$ to the balance producing the centripetal force necessary to explain the observations, yielding the equation
\begin{equation}\label{ForBal}
    F_C = F_G + F_E\,,
\end{equation}
where $F_C=mv^2/r$ is the centripetal force exerted on a test mass $m$, and $F_G=-m\frac{{\rm d}\Phi}{{\rm d}r}$ is the gravitational attraction on the same test mass $m$, exerted by the baryonic matter in the galaxy, whose gravitational potential is given by $\Phi(r)$.
What changes from one proposal to another is the form and nature of $F_E$.
For example, in the case of dark matter models, the extra force $F_E=-m\frac{{\rm d}\Phi}{{\rm d}r}$ is of gravitational nature, and its source is identified with a mass distribution, usually in the shape of a halo surrounding the galaxy.
An alternative to dark matter is the so-called Modified Newtonian Dynamics (MOND), which proposes to modify Newton's law of gravitation in the galactic regime \cite{Milgrom:1983ca,Milgrom:1983pn,Milgrom:1983zz}. 
This can also be written in the form of eq. (\ref{ForBal}).

In this work, we consider an extra force of entropic nature emerging from statistical considerations. 
Statistical physics is a suitable framework to study how macroscopic phenomena, e.g., entropic forces, may emerge in systems with a large number of interacting constituents, such as galaxies.  
Indeed, galactic constituents (stars, dust, radiation) and processes (stellar aggregation, dissipation, diffusion) provide a richer environment than the usual highly idealized considerations employed in their study \citep{Hjorth93}.
With this in mind, we investigate here the possible entropic nature of the extra force $F_E$ in eq.~(\ref{ForBal}) from the point of view of its adequacy to describe observed rotation curves.

Entropic forces are emergent phenomena occurring due to the tendency of a system to increase its entropy \citep{Verlinde:2016toy, NROOS2014, Neumann1980}.
This means that fundamental interactions take place within galaxies that aggregate to produce emergent phenomena. 
For example, in \citet{Jacobson1995, Verlinde:2016toy}, it is proposed that the very gravitational interactions are of entropic nature, and the features of the resulting equations of motion are explored. 
Similarly, in \citet{Diaz-Saldana:2018ywm}, the authors study the anomalous galactic rotation curves as emergent from an entropy-area relation derived from a Schwarzchild black-hole metric.

In a different approach to entropic forces emerging in galactic contexts, here we start with simple assumptions about the statistical system to obtain a simple model of an entropic force $F_E$.
Then, we test its adequacy to describe the observed rotation in galaxies. 
An appropriate performance of this simple model may shed light on further refinements in exploring the possible emergent nature of galactic dynamics.

Indeed, common entropic approaches to eq.~(\ref{ForBal}) assume that the whole force responsible for the centripetal pull is of entropic nature, i.e., $F_C=-T\,\nabla S$, where $S$ is the entropy; meaning that Newton's law of gravitation $F_G$ is a first-order term of the entropic force, while $F_E$ is a correction to it responsible for the discrepancies.
Since $-T\,\nabla S$ is linear in the entropy, the decomposition in eq.~(\ref{ForBal}) yields an effective decomposition of entropy $S=S_G+S_E$. 
Thus, $S_G$ should always yield Newton's law of gravitation, as any additional effect can be added to the term $S_E$.
To compare in a straightforward manner the possible entropic nature of the discrepancies with a dark matter halo model, we will focus on the fraction of the entropy responsible for the observed discrepancies $S_E$.

Our initial assumptions include that the underlying statistical system is in equilibrium, with negligible fluctuations at macroscopic scales.
We also assume a null exchange of energy or particles with the environment.
Thus, we can consider a microcanonical ensemble and a Boltzmann type of entropy.
Even more, the probability associated with each state is given as a distribution $P(r)$ over the galactic radius $r$.
Different sensible probability distributions may be investigated to possibly explain the behavior of galactic rotation curves.
For example, one could demand that $P(r)$ be a solution of an evolution equation, e.g.~Fokker-Planck equation, leading to its current distribution.

The Fokker-Planck equations have helped to study macroscopic properties of systems with many interacting particles \citep{f1914,p1917,cclr16,gqr01,h87,HFCP17}.
These equations describe the evolution of the probability density of particles subject to drag interactions, as well as random interactions. 
Regarding galaxies as a class of such systems, the working hypothesis of this investigation is that entropic forces emerge from a probability distribution that is effectively described by an F-P equation.
This article aims to explore the viability of this description.

In this context, we propose to explore a minimal entropic-force model derived from a 
simple solution of the Fokker–Planck equation (FPE). 
Compared to standard dark matter halo profiles such as NFW, Burkert, and pseudo-isothermal (ISO), 
the FPE model has the advantage of relying on a statistical–mechanical foundation with 
very few free parameters, while still retaining the ability to reproduce stationary 
configurations of large systems. 
In this work we confront the FPE model with high-quality rotation curves from the SPARC 
database, performing a systematic comparison with NFW, Burkert, and ISO fits. 
Beyond the quality of the rotation curve fits, we investigate the plausibility of the 
stellar mass-to-light ratios required by each model, and we search for possible 
correlations between the FPE parameters and the global galaxy properties. 
These tests allow us to assess whether such a minimal entropic approach can compete with 
standard halo profiles and provide new insights into scaling relations in galactic dynamics.

Beyond providing an alternative fit to galactic rotation curves, the goal of this work is to explore whether the additional force required to explain the observed dynamics can be understood as an emergent phenomenon arising from an underlying statistical description. In contrast to empirical halo profiles, which are typically introduced as phenomenological parametrizations of unseen mass distributions, the present approach is rooted in a minimal statistical–mechanical framework. In this sense, the model aims not only to reproduce the observed kinematics, but also to investigate whether global scaling relations can naturally emerge from the same underlying description. This perspective places the present work closer in spirit to other emergent approaches to gravity and galactic dynamics, while remaining agnostic about the microscopic origin of the effective statistical system.

Hence, the article is organized as follows: in section \ref{GRC}, we review the expected rotation profiles resulting from a keplerian model of galaxies and their discrepancies with the observed rotation curves.
Section \ref{EF} describes the proposal of our work. 
It presents an entropic force included in the balance of forces establishing the centripetal force that could explain the observed rotation curves.
In section \ref{DA}, we use data from the SPARC database to investigate the idea of an entropic extra force in galaxies and in section~\ref{Sec:Results}, we present the results. 
Finally, in section \ref{DFR}, we discuss our findings and provide the final remarks of our work, including prospective topics to be investigated.

\section{Galactic rotation curves}\label{GRC}

The aforementioned discrepancies in the rotation of galactic material derive from the expected notion that the centripetal pull $F_C$ corresponds only to the gravitational force $F_G$, i.~e., $F_C = F_G$, leading to the following equation:
\begin{equation}\label{NewtForBal}
    v(r) = \sqrt{ -r\frac{{\rm d}\Phi}{{\rm d}r} } \,,
\end{equation}
which attempts (yet fails) to describe observed galactic rotation curves, i.e., the tangential speed of matter around the center of the galaxy as a function of the galactic radius $r$.
Observations of the distribution of baryonic matter yield a gravitational potential $\Phi(r)\sim 1/r$ for large radii. 
However, measurements of rotation curves show that, at large radii, the decay of the angular speed is not as steep as predicted by eq. (\ref{NewtForBal}) from the presence of baryonic matter only.

In the dark matter scenario, for instance, the discrepancy is addressed by postulating a halo of weakly interacting matter surrounding the galaxy.  
This modifies the potential $\Phi$ in eq.~(\ref{NewtForBal}) and, consequently, the rotation curve. 
While different dark matter potentials have been explored \citep{baudis2016dark}, the profile suggested by Navarro, Frenk, and White (NFW) is one of the most commonly used  \citep{Navarro:1995iw}. It provides a fit for halos across a broad range of scales, from dwarf galaxies to galaxy clusters, and is consistent with the hierarchical structure formation predicted by the concordance model \citep{Cole1996,Subramanian2000}. 
However, the NFW profile suffers from the cusp problem, where the central density diverges, and, being derived from $N$-body simulations, it does not naturally incorporate baryonic feedback. 
The gravitational force from an NFW halo can be written as
\begin{eqnarray}\label{Fnfw}
     F_{\rm NFW}  &=&  
      ~G\,\frac{M_{\rm vir}\,m}{{\rm ln}(1+c)-c/(1+c)}
       \left(\frac{1}{r^2}\right)\\ 
        &&{}\quad\times\left[\frac{r}{r+R_s} - {\rm ln}\left(1+\frac{r}{R_s} \right) \right] \,,
\end{eqnarray}
where $M_{\rm vir}$, $\rho_0$, $R_s$, and $c$ are parameters that depend on the particular galaxy. 
For a gravitational pull of this sort, the corresponding mass density profile is
\begin{equation}
    \rho_{\rm NFW}(r) = \frac{\rho_0}{\left(\tfrac{r}{R_s}\right)\left(1+\tfrac{r}{R_s}\right)^2}\,.
\end{equation}

Other widely used density profiles include the nonsingular isothermal sphere (ISO) \citep{1991begeman} and the Burkert profile \citep{Burkert_1995}. 
The isothermal sphere is described by
\begin{equation}
    \rho_{\rm ISO}(r) = \frac{A}{1+\left(\tfrac{r}{b}\right)^2}\,,
\end{equation}
where $A$ and $b$ are related to the central density and a scale radius, respectively. 
This profile, while is not derived from cosmological N-body simulations, and fails to capture the correct outer behavior, has the advantage of yielding flat rotation curves at large radii and providing simple, analytically solvable Jeans equations. In addition, their velocities follow a Maxwell-Boltzmann distribution and the velocity dispersion is assumed to be isotropic and constant throughout the self-gravitating system. One of its main strengths is that it predicts flat rotation curves at large radii, consistent with observed galactic dynamics.

The Burkert density profile has the functional form \begin{equation}
    \rho_{\rm Bur}(r) = \frac{A}{\left(1+\tfrac{r}{b}\right)\left[1+\left(\tfrac{r}{b}\right)^2\right]}\,,
\end{equation}
where $A$ and $b$ have a similar meaning as in the isothermal profile and was proposed to fit dwarf and low surface brightness galaxies.
Unlike the NFW model, it avoids the central cusp and better matches observations in the inner regions, possibly due to baryonic processes such as supernova feedback flattening the core density. 

Despite their phenomenological success, both ISO and Burkert remain empirical parameterizations without clear theoretical underpinning.

In this article, we explore an alternative description of the {\it missing} force responsible for the observed galactic rotation curves, motivated not by mass distributions but by the emergence of an additional entropic force derived from statistical mechanics.

\section{Entropic forces}\label{EF}

The difference between the rotation predicted by the presence of baryonic matter in galaxies and its observed rotation may be explained by an extra term $F_E$ in the balance of radial forces: $F_C = F_G  + F_E$. Thus, the rotation curve, eq. (\ref{NewtForBal}), becomes
\begin{equation}
    v(r) = \sqrt{ r\left(-\frac{{\rm d}\Phi}{{\rm d}r} + \frac{1}{m}F_E\right)} \,.
\end{equation}
Here we consider $F_E$ as an entropic force,
\begin{equation}\label{EntFor}
    F_E = T\frac{\partial S}{\partial r}\,,
\end{equation}
where $T^{-1}=\partial S/ \partial E$ is the effective temperature parameter of an underlying statistical system.  

Entropic forces emerge from the statistical tendency of systems to maximize entropy \citep{NROOS2014}. 
In this framework, only the radial dependence of the probability distribution and its associated entropy are required to describe observed rotation curves. 
For comparison, note that the NFW profile itself can be reproduced within this entropic picture: using

\begin{equation}\label{eq:EntropyNFW}
   S(r) = \frac{A}{r}\ln \left( 1+\frac{r}{B}\right)\,.
\end{equation}
in eq.~(\ref{EntFor}) leads to the same extra force associated with the NFW potential, provided $A$ and $B$ are appropriately identified with halo parameters.

For a more fundamental starting point, consider instead a Boltzmann entropy, $S(r) = -k_B\,\ln P(r)$, where $P(r)$ is the probability distribution of microstates over the galactic radius. 
The corresponding speed profile becomes 
\begin{equation}\label{ModRot}
    v(r) = 
     \sqrt{F_G 
      - k_BT\frac{r}{m}\frac{\partial}{\partial r}{\rm ln}[P(r)]} \,,
\end{equation}
which shifts the focus from prescribing a mass density $\rho(r)$ to specifying a probability distribution $P(r)$.

As a sensible initial step, we assume galaxies are effectively isolated on observational timescales, justifying the use of a microcanonical ensemble with Boltzmann entropy. We further require that $P(r)$ be a stationary solution of the Fokker–Planck equation (FPE),
\begin{equation}\label{3dFPE}
    \frac{\partial}{\partial t} P = 
     \nabla^2\left(\epsilon P\right) 
      -\nabla\cdot\left(\vec{\mu} P \right) \,,
\end{equation}
with diffusion coefficient $\epsilon$ and drift vector $\vec{\mu}$. For stationary states, $\partial_t P=0$, yielding
\begin{equation}\label{sFPE}
    \nabla^2\left(\epsilon P\right) 
     -\nabla\cdot\left(\vec{\mu} P \right) 
      = 0 \,.
\end{equation}

\subsection*{A simple entropic force model}

To obtain a tractable case, let us consider constant $\epsilon$ (large-scale averaged diffusion) and a radial-only drift $\vec{\mu}=\mu(r)\hat{r}$ consistent with spherical symmetry. 
Equation~(\ref{sFPE}) then reduces to
\begin{equation}\label{FPsimple}
    \epsilon\,\partial_r\!\left(r^2\partial_r P\right)
     -\partial_r\!\left(r^2\mu P\right)=0\,.
\end{equation}

The drift commonly arises from a potential $U=U(r)$ and, since we are investigating galactic scenarios, where one wishes to avoid adding new interactions, let us consider that the potential is gravitationally generated. 
Integrating eq.~(\ref{FPsimple}) for a drift of the form $\mu(r)=k/r^2$, as motivated by a central Kepler potential, and imposing the boundary condition $P(r\to\infty)=0$, one obtains the radial distribution
\begin{equation}\label{eq:PPoissonLike}
 P(r) \propto  \exp\!\left(-\frac{k}{\epsilon\, r}\right) -1 \,,
\end{equation}
and the associated entropic force
\begin{equation}\label{KeplerDrift}
    F_r(r) = \frac{k/\epsilon}{\beta\,r^2\left[\exp\!\left(\tfrac{k}{\epsilon r}\right) - 1  \right]}\,,
\end{equation}
where the parameters $k/\epsilon$ and $\beta^{-1}=k_B T$ characterize each galaxy. 

\begin{figure}[ht!]
\raggedleft
\includegraphics[width=.45\textwidth]{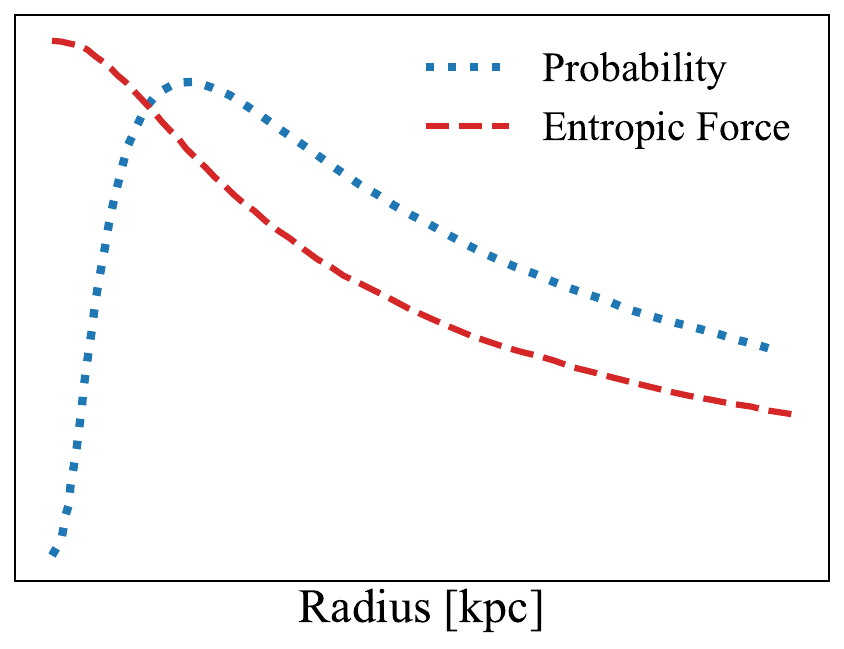}
\caption{Qualitative description of the probability distribution and its corresponding entropic force obtained from a solution of the Fokker-Planck equation with an inverse $r$-squared drift.}
\label{Kdrift}
\end{figure}

This construction shows that the FPE naturally generates an additional, radially dependent contribution to the effective force balance. 
Importantly, the resulting force tends to a constant at large radii, echoing the empirical flatness of galactic rotation curves. 
The qualitative behavior of both the probability distribution and the associated entropic force is illustrated in Fig.~\ref{Kdrift}.

At this stage, it is important to clarify the physical interpretation of the effective statistical description adopted here. The use of a Fokker–Planck equation should be understood as a coarse-grained representation of the collective dynamics of galactic constituents, rather than as a literal description of microscopic particle diffusion. In this context, the diffusion coefficient $\epsilon$ and the effective temperature $T$ parametrize the net effect of complex processes such as gravitational interactions, scattering, phase mixing, and dynamical relaxation. From this perspective, the resulting probability distribution $P(r)$ encodes the stationary configuration of an effective ensemble, and the associated entropic force captures the macroscopic response of the system to these underlying processes. This interpretation allows us to remain agnostic about the detailed microphysics while still providing a well-defined and testable framework at galactic scales.

\section{Data analysis}\label{DA}

\begin{figure}[ht!]
\raggedright
\includegraphics[width=.5\textwidth]{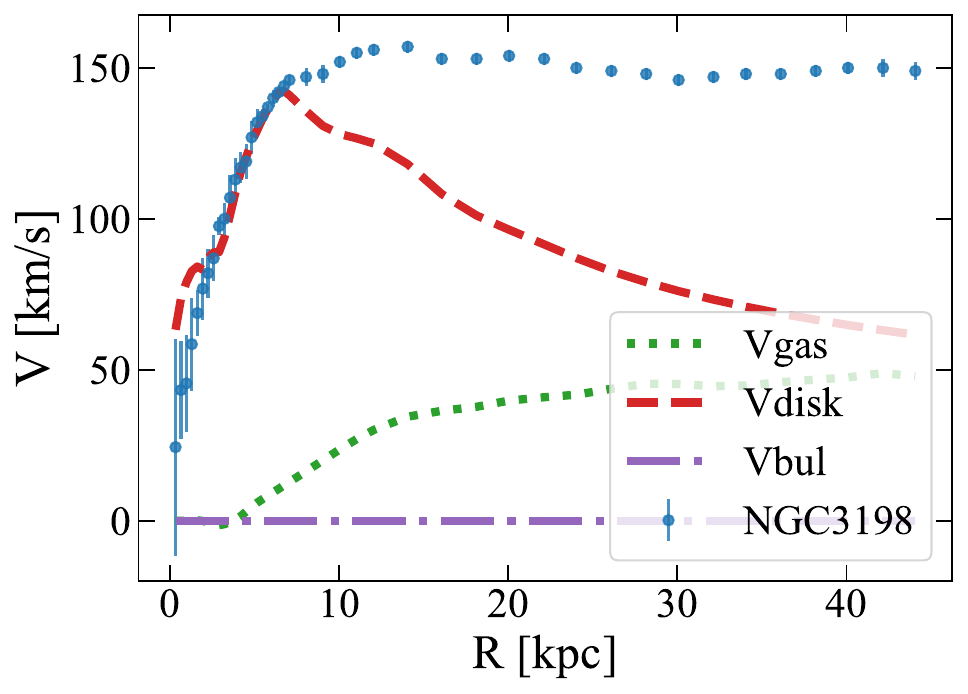}
\caption{NGC3198 rotation curve (blue points). 
Also shown is the stellar disk (red dashed line), the gas contribution (green dotted line), and bulge one (dotted dashed line) null in this case,  as given in~\citet{Lelli:2016zqa}} 
\label{fig:RotCurvProb}
\end{figure}

To test the model introduced in the previous section, let us compare it to rotation curves compiled in the SPARC project.
The SPARC database is a sample of 175 nearby galaxies with surface photometry at 3.6 $\mu$m and  high-quality rotation curves from previous $H_I/H_\alpha$ studies~\citep{Lelli:2016zqa}.
SPARC spans a broad range of morphologies (S0 to Irr), luminosities ($\sim 5$ dex), and surface brightnesses ($\sim 4$ dex).
Hereafter, capital $(V,R)$ will refer to data.
Figure \ref{fig:RotCurvProb} shows a typical galactic rotation curve as observed, as well as the contributions expected from each type of detectable (baryonic) matter.

In order to proceed, we define the variable 
\begin{equation}\label{eq:AV}
\Delta V^2 \equiv V^2_{\rm observed} - (V^2_{\rm gas} + \Upsilon \, V^2_{\rm star}),
\end{equation}
where $\Upsilon=M^*/L$ is the stellar mass-to-light ratio, $V_{\rm obs}(R)$ the observed velocity, $V_{\rm gas}$ the contributions from gas, and $V^2_{\rm star}$ includes the $V_{\rm disk}$ the stellar disk , and $V_{\rm bulge}$ stellar bulge.

Eq.~(\ref{eq:AV}) accounts for the discrepancy between the (squared) observed rotation speed ($\Delta V^2$) and the expected one from the baryonic matter content in each galaxy~\citep{klmdbs16}.
Using eq.~(\ref{ModRot}) for entropic forces, eq.~(\ref{eq:AV}) becomes
\begin{equation}\label{eq:AV2}
    \Delta V^2 = 
    -\frac{r}{\beta m}\frac{{\rm d}}{{\rm d}r}{\rm ln}P(r) .
\end{equation}
We can note that for any given sensible choice of diffusion $\epsilon$ and drift $\mu$ in eq.~(\ref{sFPE}), 
we can obtain a probability function $P(r)$ that, in turn, yields a rotation curve profile to be compared against observed data.
Two working assumptions are considered in what follows: the stellar mass-to-light ratio $\Upsilon$ and the test mass $m$ are considered constant for all radii in each rotation curve.
Now, let us consider the simple probability distributions described in the previous section.

For the case of an extra force given by~(\ref{KeplerDrift}), corresponding to a simple {\it Fokker-Planck entropic} (FPE) force model, the discrepancy is
\begin{equation}\label{eq:M2}
     \Delta V^2_{\rm FPE} = \frac{k/\epsilon}
     {\beta\,m\,r\left[{\rm exp}\left(k/\epsilon\, r\right) - 1  \right]}\,
     = \frac{ab}{r}
     {\left[{\rm exp}\left(b\, /r\right) - 1  \right]}^{-1},
\end{equation}
where $a^{-1}=\beta m$ and $b=k/\epsilon$. For comparison with standard halo profiles, an equivalent effective density can be defined via $\Delta V^2 = G M_{\rm eff}(r)/r$, leading to $\rho_{\rm FPE}(r)=\frac{a\,b^2}{4\pi G\,r^4}\frac{e^{b/r}}{(e^{b/r}-1)^2}$.
While for the case of the NFW, Isothermal, and Burkert, models, the squared discrepancies are given the following expressions, respectively:
\begin{equation}\label{eq:M3}
     \Delta V^2_{\rm NFW} = 
      \frac{A}{m}
       \left(\frac{1}{r}\right)
        \left[\frac{r}{r+B} - {\rm ln}\left(1+\frac{r}{B} \right) \right] \,.
\end{equation}
\begin{equation}\label{eq:M4}
     \Delta V^2_{\rm Iso} = 
      A B^2 - \frac{1}{r}\left[A B^3 \tan ^{-1}\left(\frac{r}{B}\right) + 1  \right] \,.
\end{equation}
\begin{equation}\label{eq:M5}
     \Delta V^2_{\rm Bur} =\frac{AB^3}{r}\left[{\rm ln}\left[\left(B^2+r^2\right)\left(B+r\right)^2\right]-2 \tan ^{-1}\left(\frac{r}{B}\right)
\right]\;,
\end{equation}
being $A$ and $B$ the parameters for each model to be fitted.
Using the SPARC data set, we select edge-on rotation curves of galaxies with an inclination greater than $30^\circ$~\cite{Lelli:2016zqa}, quality flag at least medium (1 and 2), and  at least four data points in its rotation curve. 
This yields a set of 93 galactic rotation curves. 

For each galaxy we used the radial profiles of the observed velocity $V_{\rm obs}(R)$, its uncertainty $\sigma_{V_{\rm obs}}$, and the contributions from gas ($V_{\rm gas}$), and stars , taken as the addition of the stellar disk ($V_{\rm disk}$) and stellar bulge ($V_{\rm bulge}$) ($V_{\rm star}^2=V_{\rm disk}^2+V_{\rm bulge}^2$). We worked in squared velocities, such that the observational quantity is $V_{\rm obs}^2$.

In all cases the predicted squared velocity was written as
\begin{equation}
V_{\rm model}^2 (R) = \Delta V^2(R) + V_{\rm gas}^2(R) + \Upsilon \, V_{\rm star}^2(R),
\end{equation}
where $\Delta V^2(R)$ corresponds to the dark matter (or effective) contribution described in Sec.~II, and $\Upsilon$ is the stellar mass-to-light ratio in the 3.6~$\mu$m band. This parameter was treated as a free variable in all fits. 

We performed non-linear least-squares fits using the \texttt{lmfit} Python package. The minimized statistic was
\begin{equation}
\chi^2 = \sum_i \frac{\big(V_{\rm obs}^2(R_i) - V_{\rm model}^2(R_i)\big)^2}{\sigma_{V^2}(R_i)^2},
\end{equation}
with the reduced chi-square defined as $\chi^2_{\rm red} = \chi^2 / N_{\rm dof}$. We imposed physically motivated priors: all scale parameters were constrained to be positive, and the stellar mass-to-light ratio was restricted to the interval $0.15 \leq \Upsilon \leq 2.5$, consistent enough with stellar population synthesis estimates for the SPARC sample.
%
For each fit we recorded the best-fit parameters, their uncertainties, the reduced chi-square, and the Akaike and Bayesian information criteria (AIC and BIC). These quantities allow both an absolute assessment of fit quality and a relative comparison across different models. We then analyzed the distributions of these statistics across the full sample using boxplots and cumulative distribution functions (CDFs).

It is worth emphasizing that, unlike standard halo profiles such as the pseudo-isothermal or Burkert models, which are primarily empirical parametrizations of the mass distribution, the FPE model is not introduced as a density profile but rather as the outcome of an underlying statistical framework. In this sense, the functional form of the additional contribution to the rotation curve is not imposed a priori, but emerges from the stationary solution of the Fokker–Planck equation under simple and physically motivated assumptions. This distinction is important when interpreting the results, as it suggests that the success of the model may be pointing to an underlying statistical origin of the observed dynamics, rather than merely providing an alternative fitting function.


\section{Results}\label{Sec:Results}
We applied the fitting procedure described in Sec.~III to the SPARC sample.

\begin{figure}[h]
\centering
\includegraphics[width=0.48\textwidth]{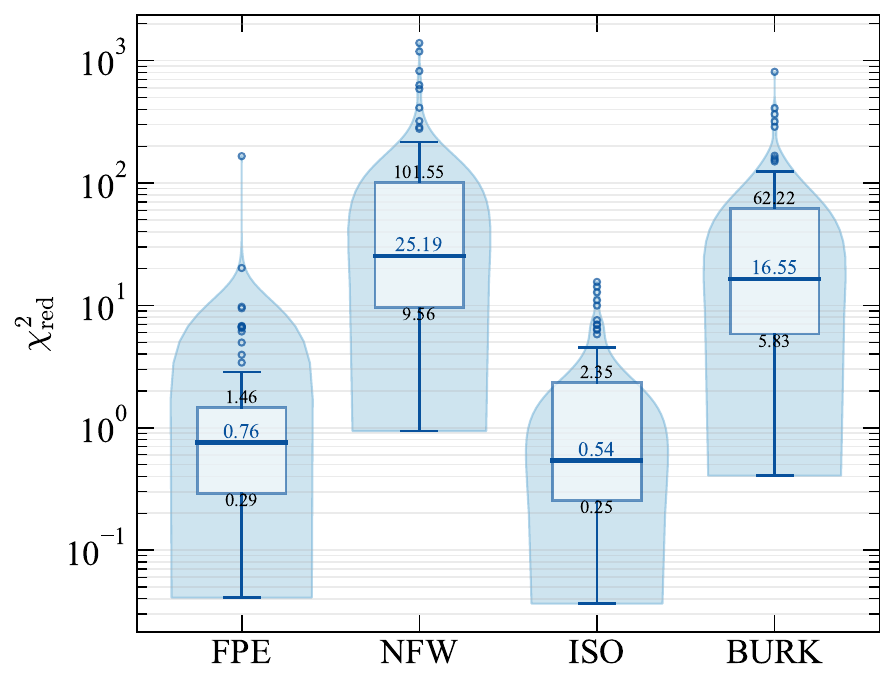}
\caption{Distributions of reduced chi-square $\chi^2_{\rm red}$ for the four models. 
Each violín represents the probability density of the distribution, with an internal box showing 
the first quartile, median (annotated), and third quartile. 
FPE and ISO are sharply peaked at $\chi^2_{\rm red}<1$, while NFW and Burkert display broad 
tails extending to $\chi^2_{\rm red}\gg 1$.}
\label{fig:chi2_violin}
\end{figure}

Figure~\ref{fig:chi2_violin} shows the distributions of reduced chi-square values for the four models, 
using violin plots overlaid with box statistics (quartiles and medians). 
The violin shapes correspond to kernel density estimates, providing a smooth 
representation of the distribution of fit results across galaxies. 
The FPE and ISO profiles are strongly concentrated at $\chi^2_{\rm red}\lesssim 1$, 
whereas NFW and Burkert exhibit broad distributions extending up to two orders of magnitude higher. 
This confirms the systematic statistical advantage of the FPE and ISO models.
 
\begin{figure}[h]
\centering
\includegraphics[width=0.48\textwidth]{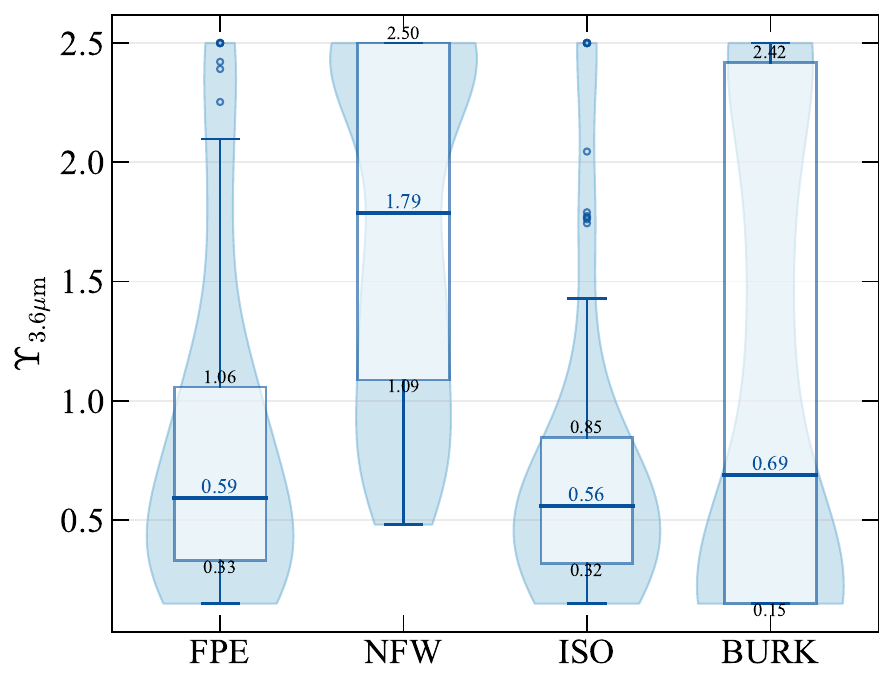}
\caption{Distributions of the stellar mass-to-light ratio $\Upsilon_{3.6\mu{\rm m}}$ obtained from 
the fits. The violin plots show kernel density estimates of the distribution, 
while the internal boxes mark quartiles and medians (annotated). 
FPE and ISO yield values consistent with stellar population estimates 
($0.3\lesssim\Upsilon\lesssim1.0$), whereas NFW and Burkert tend to saturate the upper bound 
of the prior, reflecting tension with physically plausible stellar contributions.}
\label{fig:upsilon_violin}
\end{figure}

Figure~\ref{fig:upsilon_violin} displays the distributions of the stellar mass-to-light ratio $\Upsilon_{3.6\mu{\rm m}}$ 
obtained in the fits. 
Here again the violin shapes represent kernel density estimates, while the boxes mark quartiles and 
medians. 
FPE and ISO yield physically consistent values in the range $0.3\lesssim\Upsilon\lesssim1.0$, 
while NFW and Burkert frequently drive $\Upsilon$ to the upper or lower bound of the prior interval. 
This highlights the difficulty of the classical halo profiles in reproducing the observed curves 
without invoking unrealistically large stellar contributions.

\begin{table*}[ht!]
\small
\centering
\caption{Median fit statistics for the four models across the SPARC sample. 
Values shown are the medians of the reduced chi-square ($\chi^2_{\rm red}$), 
Akaike Information Criterion (AIC), Bayesian Information Criterion (BIC), 
and stellar mass-to-light ratio $\Upsilon_{3.6\mu{\rm m}}$. 
The last column reports the fraction of galaxies in which the fit reached 
the upper bound of the $\Upsilon$ prior ($\Upsilon=2.5$).}
\label{tab:global_stats}
\begin{tabular}{lccccc}
\hline\hline
Model & \quad Median $\chi^2_{\rm red}$ & \quad Median AIC & \quad Median BIC & \quad  Median $\Upsilon$ & \quad \%$\Upsilon_{\geq2.5}$ \\
\hline
FPE   & 0.76 & $-1.95$ & $-0.21$ & 0.59 & 17.2 \\
ISO   & 0.54 & $-4.80$ & $-3.90$ & 0.56 & 13.98 \\
NFW   & 25.19 & 62.05  & 64.89   & 1.79 & 56.99 \\
BURK  & 16.55 & 52.28  & 54.97   & 0.69 & 36.56 \\
\hline
\end{tabular}
\end{table*} 

This trend is also reflected in the information criteria and stellar mass-to-light ratios. 
Table~\ref{tab:global_stats} summarizes the median fit statistics across the sample. 
FPE and ISO consistently achieve median reduced chi-square values below unity, 
together with significantly lower AIC and BIC scores compared to NFW and Burkert. 
In contrast, NFW and Burkert yield median $\chi^2_{\rm red}$ values of $\sim25$ and $\sim17$, 
respectively, and much larger information criteria. 
The table also shows that while the median stellar mass-to-light ratios for FPE and ISO 
lie within the physically expected range $0.5\lesssim \Upsilon \lesssim 1.0$, 
a substantial fraction of NFW (57\%) and Burkert ($\sim$37\%) fits are pushed to the 
upper prior bound $\Upsilon=2.5$, indicating that the classical halo profiles 
require implausibly high stellar contributions in order to reproduce the observed curves.

\begin{figure*}[ht!]
\centering
\includegraphics[width=0.8\textwidth]{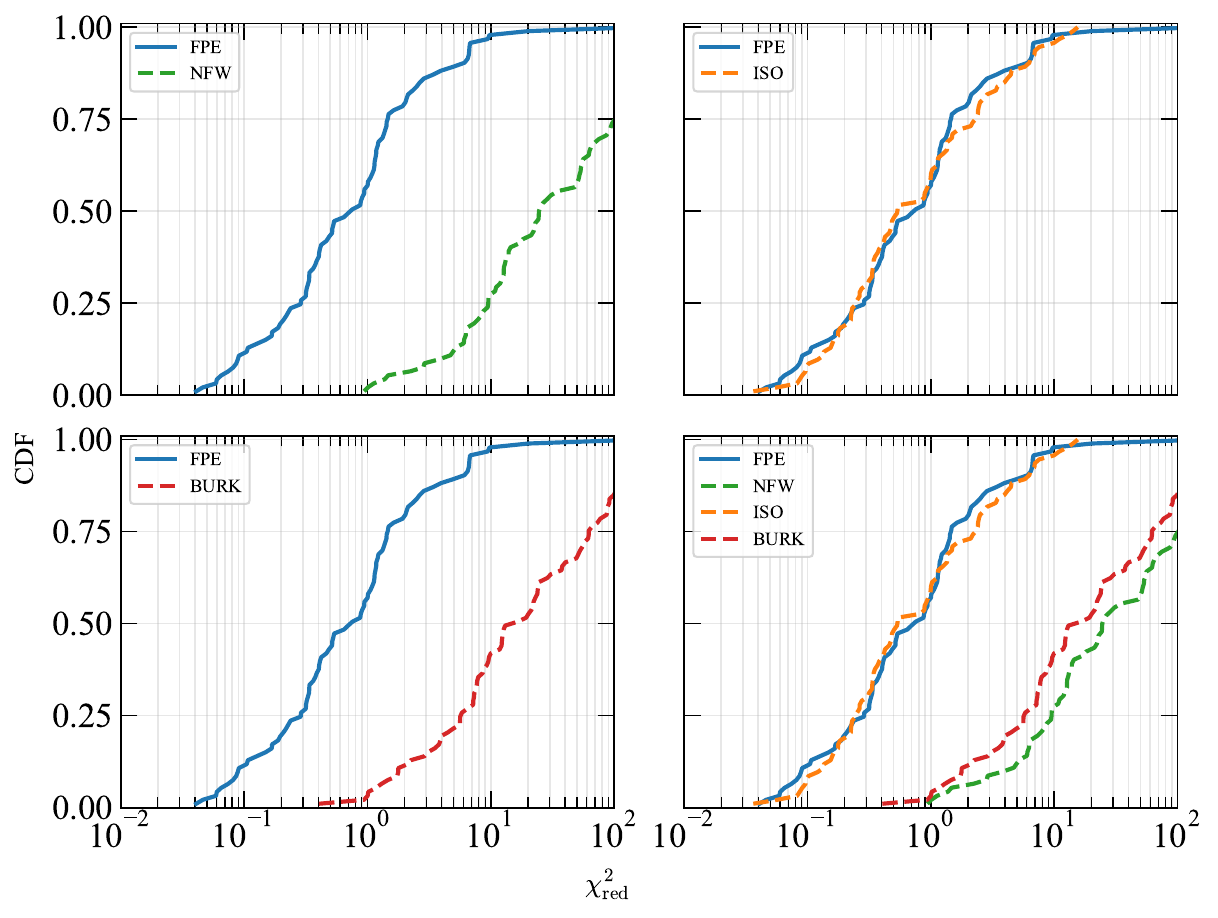}
\caption{Cumulative distribution functions (CDFs) of the reduced chi-square $\chi^2_{\rm red}$ 
for the four models. 
The steep rise of the FPE and ISO curves indicates that most galaxies reach 
$\chi^2_{\rm red}<1$, whereas NFW and Burkert accumulate more slowly, 
with a significant fraction of cases extending to $\chi^2_{\rm red}\gg 1$. 
}
\label{fig:CDF}
\end{figure*}

Finally, we analyzed the distributions using cumulative distribution functions (CDFs), 
which provide a global view of the fraction of galaxies below a given $\chi^2_{\rm red}$. 
As shown in Fig.\ref{fig:CDF}, the CDFs for FPE and ISO rise steeply, with most galaxies satisfying 
$\chi^2_{\rm red}<1$, while the NFW and Burkert distributions remain broad, 
extending to $\chi^2_{\rm red}\gg 1$. 
This reinforces the systematic advantage of the FPE and ISO profiles observed in the 
previous figures.

We also tested robustness under different radial cuts, excluding the innermost $R<2,3,4$~kpc. 
The qualitative hierarchy of models remains unchanged for FPE regardless the choice of $R_{\min}$.

\begin{figure*}[]
\centering
\begin{tabular}{cc}
\includegraphics[width=0.45\textwidth]{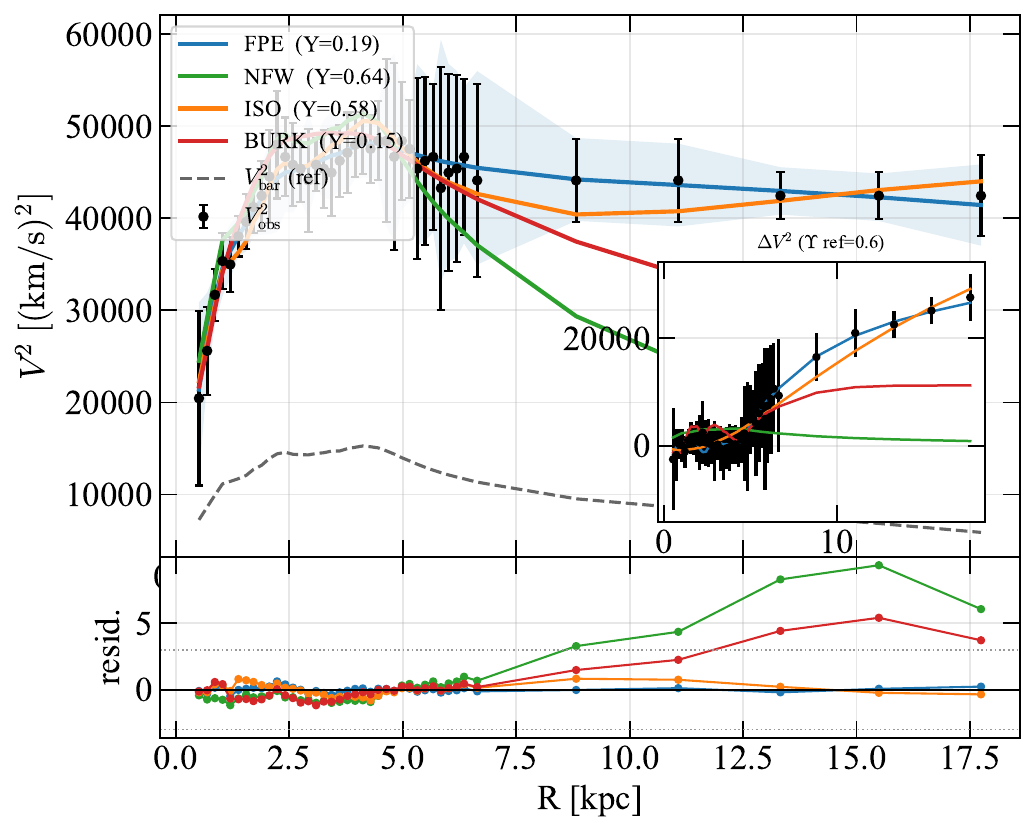} &
\includegraphics[width=0.45\textwidth]{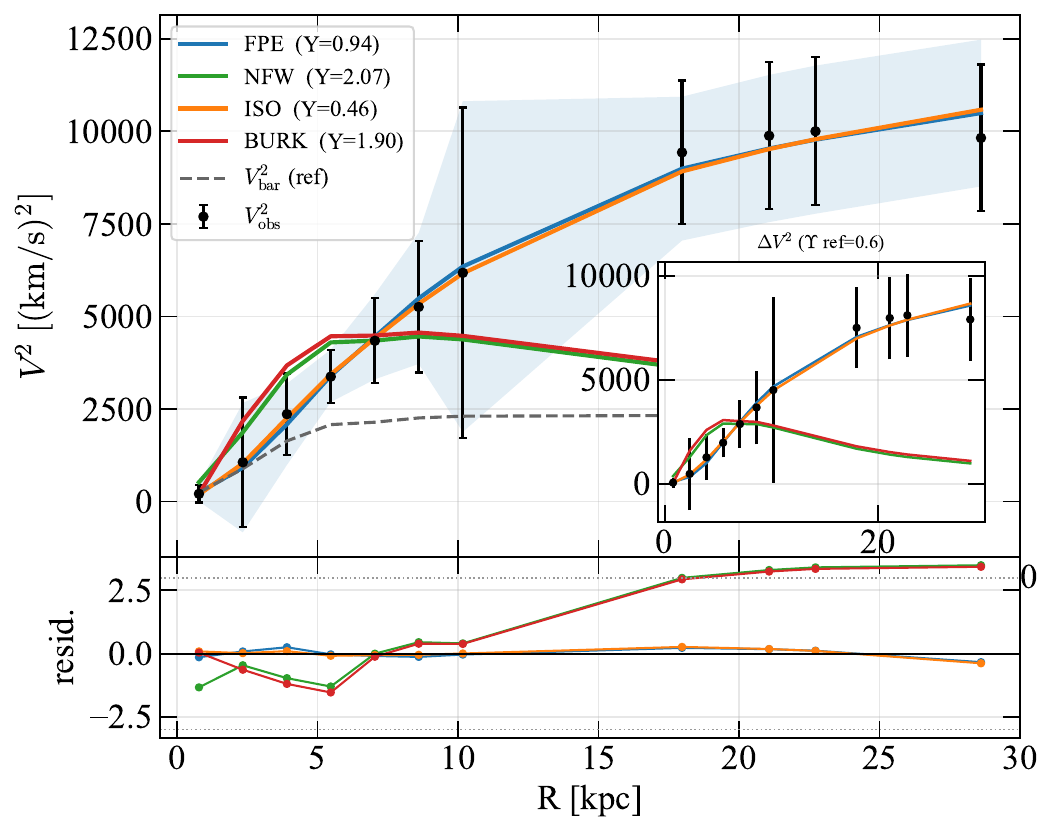} \\
(a) NGC3521 & (b) UGC05005 \\
\includegraphics[width=0.45\textwidth]{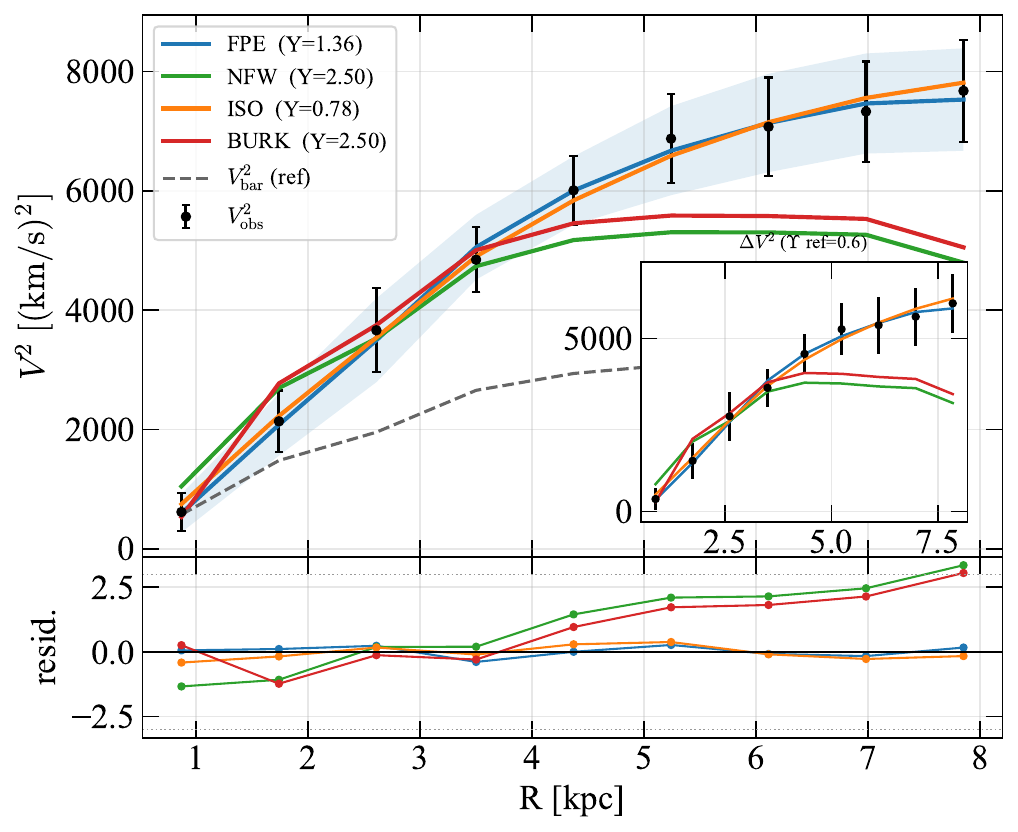} &
\includegraphics[width=0.45\textwidth]{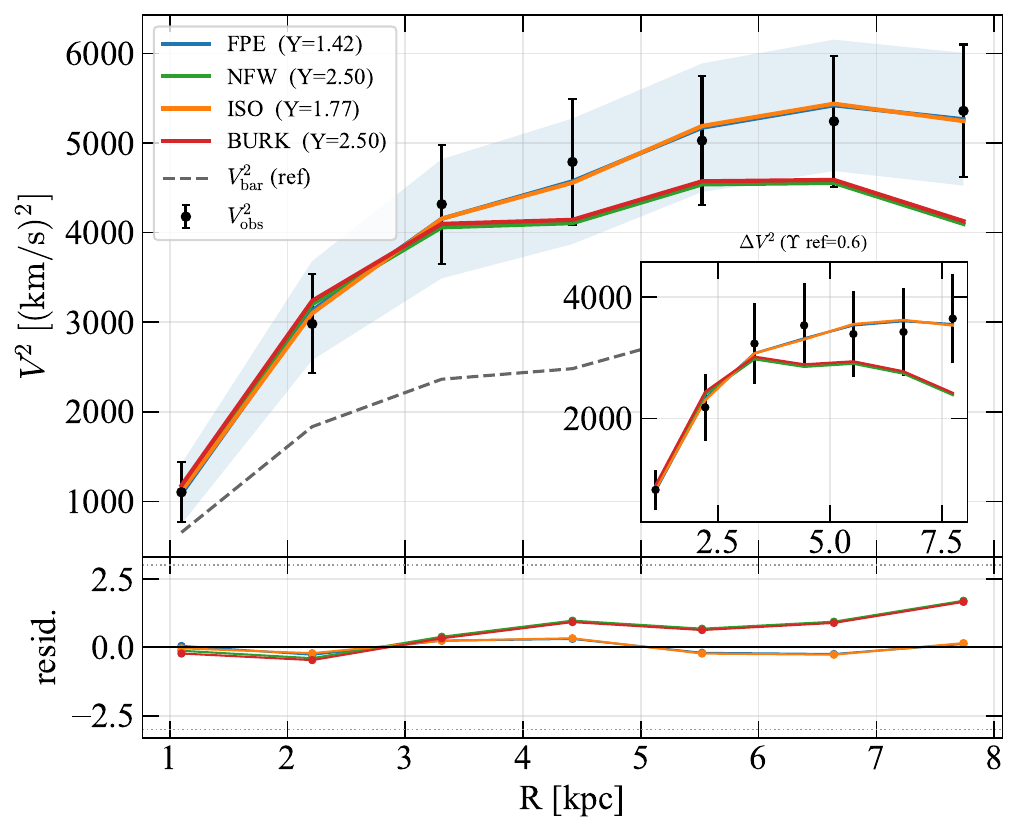} \\
(c) UGC06399 & (d) UGC10310 \\
\end{tabular}
\caption{Representative galaxies illustrating different behaviors in the fits. 
(a) NGC3521: all four models provide acceptable fits with $\Upsilon < 1$. 
(b) UGC05005: FPE and ISO yield good fits with $\Upsilon \sim 1$, while NFW and Burkert 
require $\Upsilon \sim 2$. 
(c) UGC06399 and (d) UGC10310: in both cases FPE and ISO converge to consistent 
$\Upsilon < 1.8$, while NFW and Burkert saturate the prior upper bound $\Upsilon=2.5$. Each case includes a comparison to a reference $\Upsilon=0.6$ for all models.
Together, these examples highlight the robustness of FPE and ISO and the tendency 
of classical halos to require unrealistically high stellar contributions.}
\label{fig:examples}
\end{figure*}

\paragraph{Representative galaxies}
To illustrate individual cases we present four galaxies that highlight different situations 
within the global trends. 
Figure~\ref{fig:examples}a shows NGC3521, a system where all four models provide acceptable fits 
with $\Upsilon < 1$, demonstrating that FPE can reproduce rotation curves without degrading 
cases already well described by classical profiles. 
Figure~\ref{fig:examples}b presents UGC05005, where FPE and ISO yield good fits with 
$\Upsilon \sim 1$, but NFW and Burkert require values around $\Upsilon \sim 2$ to approach 
the data. 
Figures~\ref{fig:examples}c and~\ref{fig:examples}d display UGC06399 and UGC10310, respectively, 
both of which show that while FPE and ISO converge to physically more consistent 
$\Upsilon < 1.8$, the NFW and Burkert fits are pushed to the upper prior bound 
($\Upsilon=2.5$). 
These examples reinforce the statistical preference for FPE and ISO observed in the 
global distributions, and highlight the tendency of classical halos to demand 
implausibly large stellar contributions.

\bigskip

Once the galactic rotation curves are fitted, we investigate possible relations between the FPE parameter $a$ and global properties of galaxies provided by the SPARC database. Among several combinations tested, two correlations stand out: $a$ versus the velocity along the flat part of the rotation curve ($V_{\rm f}$), and $a$ versus the total stellar luminosity at $3.6\,\mu{\rm m}$, $L_{[3.6]}$.

For the $a$–$V_{\rm f}$ relation we excluded systems without a well-defined flat velocity ($V_{\rm f}=0$). In the remaining 83 galaxies, a strong monotonic correlation is found in logarithmic scales. The linear regression yields
$\log V_{\rm f} = (0.472 \pm 0.003)\,\log a + (0.134 \pm 0.014)$,
with a Pearson correlation coefficient of $r=0.96$ ($p\simeq 10^{-46}$) and a Spearman coefficient of $\rho=0.97$ ($p\simeq 10^{-53}$). This tight correlation is natural, since the FPE model can be expressed in the form $V^2 \sim c_1 a + c_0$, linking directly the fitted parameter $a$ with the characteristic rotational velocity of the galaxy.

In the case of $a$–$L_{[3.6]}$, we identified three outliers (F568-3, F583-4, and UGC05750) with anomalously small values of $a$ ($\log a<-5$) that distorted the distribution. After excluding them, 90 galaxies remain, showing a robust linear trend in logarithmic scales:
$\log L_{[3.6]} = (1.63 \pm 0.12)\,\log a + (-5.39 \pm 0.54)$,
with $r=0.91$ ($p\simeq 10^{-34}$) and $\rho=0.92$ ($p\simeq 10^{-38}$).

NGC~5985 appears as the rightmost point in this relation due to its unusually large value of $a$. When this galaxy is excluded, the correlation becomes
$\log L_{[3.6]} = (1.78 \pm 0.12)\,\log a + (-6.05 \pm 0.53)$,
with $r=0.92$ ($p\simeq 10^{-38}$) and $\rho=0.92$ ($p\simeq 10^{-38}$).

The unusually large value of $a$ found for NGC~5985 is plausibly linked to its location at the high-velocity end of the sample. Since the FPE parameter $a$ scales approximately as $V_f^2$, galaxies with particularly large flat rotation velocities are naturally expected to populate the upper end of the $a$ distribution. Within the FPE framework, $a$ effectively controls the strength of the additional entropic contribution to the radial force balance. Therefore, galaxies requiring stronger support at large radii are expected to yield larger fitted values of $a$. In this sense, NGC~5985 appears to represent an extreme but physically consistent system rather than a pathological outlier, reflecting its position at the massive, high-$V_f$ end of the sample.

In general, these correlations reinforces the interpretation of $a$ as a parameter tied to the global baryonic content: since $a$ is proportional to the effective temperature scale of the entropic system, the observed trend links stellar luminosity with the underlying potential traced by the FPE fits.

In addition, for $\log a$–$\log L_{[3.6]}$, the linear fit provides a very good description ($r=0.92$). A quadratic form yields a marginally lower AIC but does not improve BIC nor the statistical significance of the coefficients, so we adopt the linear model as the most robust representation.

\begin{figure*}
\centering
\includegraphics[width=0.45\textwidth]{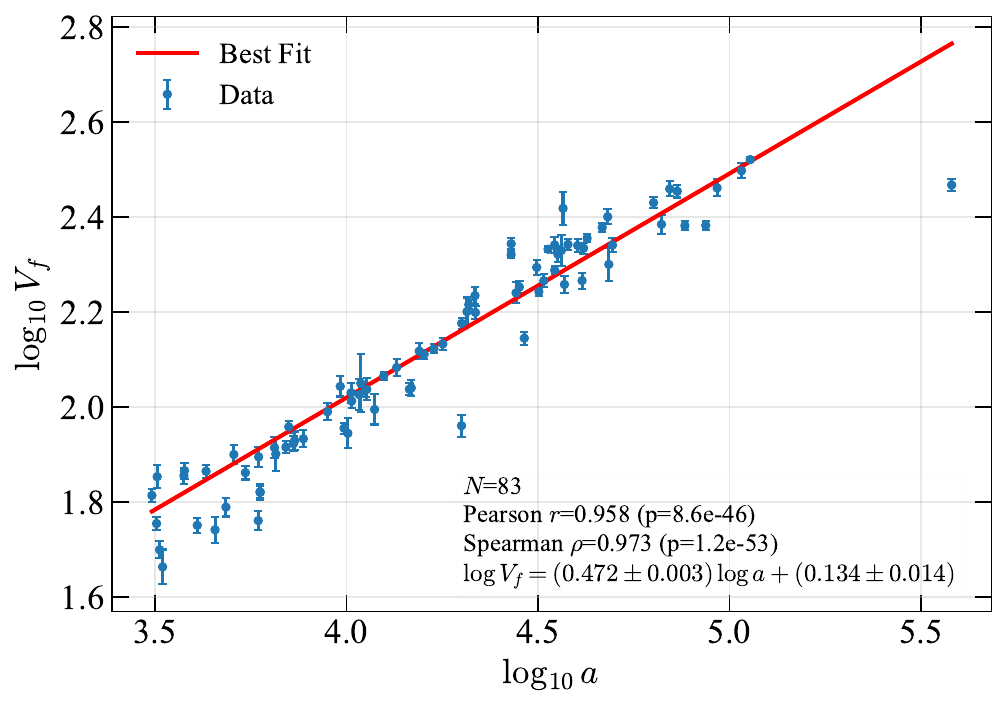}
\includegraphics[width=0.45\textwidth]{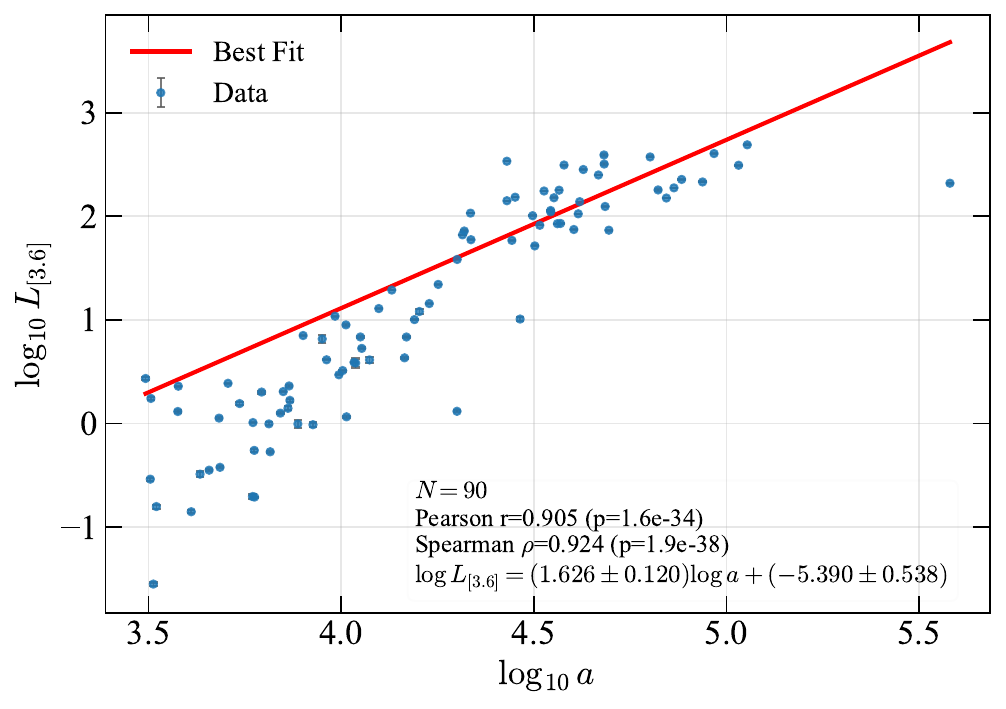}
\caption{Correlations between the FPE parameter $a$ and global galaxy properties from the SPARC sample. 
Panel (\textbf{a}) shows the relation with the flat rotation velocity $V_{\rm f}$, 
and panel (\textbf{b}) with the stellar luminosity at $3.6\,\mu$m, $L_{[36]}$. 
Outliers with anomalous values have been excluded as discussed in the text.}
\label{Fig:Characteristics}
\end{figure*}

Figure~\ref{Fig:Characteristics} displays these correlations. Panel (a) shows the $\log V_{\rm f}$–$\log a$ relation, while panel (b) presents the $\log L_{[3.6]}$–$\log a$ trend after outlier removal.

An interesting aspect of these correlations is that they naturally connect the FPE parameter $a$ with well-known galaxy scaling relations. 
Combining the relations $\log V_f \propto \log a$ and $\log L_{[3.6]} \propto \log a$ obtained here implies approximately $L_{[3.6]} \propto V_f^{3.5}$. 
This scaling is remarkably close to the classical Tully–Fisher relation, which links the luminosity of spiral galaxies with their rotational velocity as $L \propto V_f^{3-4}$. 
In this sense, the parameter $a$ emerging from the FPE description appears to encode information related to the global dynamical state of galaxies. 
Rather than being an arbitrary fitting parameter, it may capture aspects of the same underlying physics that gives rise to the empirical scaling relations observed in disk galaxies.

In this context, the parameter $a$ acquires a particularly interesting interpretation. Beyond being a fitting parameter, it effectively controls the strength of the entropic contribution to the radial force balance and appears to be directly linked to the dynamical scale of the system. Its tight correlation with both $V_f$ and $L[3.6]$ suggests that $a$ encapsulates information about the global baryonic and kinematic state of galaxies. From the statistical perspective adopted here, this behavior is consistent with interpreting $a$ as an emergent parameter associated with the effective temperature or energy scale of the underlying ensemble, providing a possible bridge between microscopic statistical properties and macroscopic observables.

\section{Discussion and final remarks}\label{DFR}

Although the model was introduced as a phenomenological description of the additional force required to explain galactic rotation curves, its structure admits a natural statistical interpretation. 
In particular, the stationary probability distribution obtained from the Fokker--Planck equation,
\[
P(r)\propto \exp\!\left(-\frac{k}{\epsilon r}\right)-1,
\]
closely resembles the structure of a Boltzmann distribution in a central potential. 
Equilibrium statistical distributions typically scale as $P(r)\propto e^{-\beta V(r)}$, and a gravitational potential behaves as $V(r)\propto 1/r$. 
In the present case, the additional ``$-1$'' term arises from the specific stationary solution of the radial Fokker--Planck equation under the imposed boundary condition. 
This suggests that the entropic contribution to the radial force balance may be interpreted as the emergent response of an effective statistical system subject to a central gravitational drift and diffusion.

Within this framework, the additional force derived from the entropy gradient provides an effective contribution to the rotation dynamics of galaxies. 
A notable feature of the resulting force is that it approaches a constant value at sufficiently large radii. 
This behavior naturally leads to approximately flat rotation curves at large radii,
one of the key observational properties of disk galaxies.
In this sense, the entropic formulation captures one of the key phenomenological requirements traditionally attributed to dark matter halos. The model therefore offers an alternative perspective to the standard dark matter interpretation of galactic dynamics. 

It is important to stress that the present framework should be interpreted as an effective description of galactic dynamics at the level of rotation curves, rather than as a complete alternative to the dark matter paradigm. While the model successfully reproduces the observed kinematics and captures key scaling relations, it does not, in its current form, address other phenomena commonly attributed to dark matter, such as gravitational lensing, large-scale structure formation, or cosmic microwave background anisotropies. In this sense, the entropic-force contribution explored here may be understood either as an effective emergent component within galaxies or as a complementary description that captures aspects of the dynamics not explicitly encoded in standard approaches. Clarifying this connection remains an open question for future work.

To test the idea of an entropic force contributing to the centripetal pull in galaxies, we fitted the observations in the SPARC database with a model that adds an entropic force to the one expected from the baryonic mass in each galaxy. 
Results show that the rotation curves in the sample have strong compatibility with the FPE model, with typical stellar mass-to-light ratios of $\Upsilon \sim 0.6$, in agreement with values reported in the literature.

For comparison, the same tests were conducted for the NFW, Burkert, and pseudo-isothermal (ISO) models. 
While ISO yields acceptable fits in several cases, both NFW and Burkert often require stellar mass-to-light ratios saturating the prior upper bound, in contrast with the more physically consistent values obtained for the FPE model. 
This highlights an appealing feature of the entropic-force description: it reproduces the observed curves without demanding implausibly high stellar contributions. 
Moreover, the minimal nature of the FPE framework leaves room for less restrictive conditions on the underlying Fokker–Planck dynamics, which could in turn describe the observed rotation curves even more accurately.

If the FPE model indeed governs the probability distribution underlying the entropic force, it implies that diffusion and drag processes effectively yield a departure from the Newtonian expectation for rotation curves. 
Furthermore, since the Fokker–Planck equation describes the time evolution of a probability density, it suggests that galaxies might undergo a transient stage that ends in the stationary configuration analyzed here. Exploring the nature of such a transient stage is a promising avenue for future work.

Entropic forces, being emergent statistical phenomena, describe the cumulative behavior of a large number of microscopic interactions. 
It is not yet clear what the fundamental mechanism behind entropic forces in galaxies might be, although several authors have investigated this problem~\cite{Verlinde:2016toy,Jacobson1995,Diaz-Saldana:2018ywm}. 
Further progress in this direction could shed light on the meaning of the effective temperature parameter and its connection with global galaxy properties.

The correlations uncovered between the FPE parameter $a$ and the global galaxy properties $V_f$ and $L_{[3.6]}$ are also remarkable. 
They echo well-established empirical laws such as the Tully–Fisher relation (luminosity–rotation velocity) and, by analogy, the Faber–Jackson relation (velocity dispersion–luminosity for ellipticals). 
Interestingly, the observed scaling $a \propto V_f^2$ suggests that the FPE parameter traces the dynamical velocity scale of galaxies. 
In this context, $a$ appears as an emergent parameter that encapsulates the scaling between baryonic content and kinematic support.

Interestingly, the $a$–$L_{[3.6]}$ correlation loosely resembles the form of a ``main sequence'' relation, reminiscent of the Hertzsprung–Russell diagram, although emerging here at the galactic rather than stellar scale. We interpret this as a heuristic analogy pointing to the thermodynamic-like role of $a$.

Also, as described here, an entropic extra force constitutes an alternative to the popular dark matter models. 
Whereas dark matter models invoke weakly interacting mass distributions in galaxies, entropic forces imply the existence of statistical systems with states described by probability distributions within galaxies. 
Nonetheless, dark matter is also invoked to explain other phenomena in astrophysics and cosmology, such as gravitational lensing, large-scale structure, or the anisotropies of the cosmic microwave background. 
In this sense, entropic forces have been used to study problems beyond galactic rotation curves, including the evolution of the universe \citep{Komatsu2014}, possible links to the dark energy problem \citep{Basilakos2014}, or scenarios in brane cosmology \citep{Ling_2010}. 
Exploring these broader implications in the specific context of the FPE framework remains open for future work.

We envision future investigations in two immediate directions. 
On the one hand, exploring alternative formulations of entropic forces, such as more refined solutions of the Fokker–Planck equation or other sensible probability distributions and entropy functions, may lead to improved descriptions of rotation curves. 
On the other hand, understanding the fundamental system from which the entropic force emerges, and assessing whether such a framework can address other open problems in astrophysics and cosmology, are key challenges ahead. 
There are already examples in the literature along these lines \citep{CCGMT17, CCGT18}, and we hope that the present work and its extensions can contribute to such efforts.

Overall, our results show that a minimal entropic-force inspired model can compete with, and in some cases outperform, standard halo profiles in describing galactic rotation curves, while naturally producing scaling relations with global galaxy properties. 
This dual success—reproducing rotation curves while revealing scaling relations with global galaxy properties—makes the FPE framework a promising avenue for further exploration.

\begin{acknowledgments}
We thank A. Hernández-Almada, A. Molgado, V. de Souza, and A. Su\'arez, for helpful comments and discussions. PIR-B acknowledge the support from the Consejo Nacional de Ciencia y Tecnología (CONACyT), México, {\it Estancias posdoctorales por M\'exico 2021}, grant No. 1233881. H.M-H. acknowledges support by SECIHTI CBF2023-2024-1630.
\end{acknowledgments}

\bibliography{bibliography}

\end{document}